\documentclass[12pt,a4paper,twoside]{article}
\usepackage[british]{babel}
\usepackage{epsf}
\usepackage{dcolumn}
\usepackage{rotate}

\newcommand{\gev}{\ensuremath{\mathrm{GeV}}}
\newcommand{\gevc}{\ensuremath{{\mathrm{GeV}\!/c}}}
\newcommand{\gevcc}{\ensuremath{{\mathrm{GeV}\!/c^2}}}

\newcommand{\nb}{\ensuremath{\mathrm{nb}}}       
       
\newcommand{\pb}{\ensuremath{\mathrm{pb}}}       
\newcommand{\invpb}{\ensuremath{{\mathrm{pb}^{-1}}}}

\newcommand{\roots}{\ensuremath{{\sqrt{s}}}}

\newcommand{\syst}{({\rm syst.})}
\newcommand{\stat}{({\rm stat.})}

\newcommand{\nxsec}{0.64}
\newcommand{\nxsstat}{\ensuremath{\mbox{}^{0.12}_{0.11}}}
\newcommand{\nxssys}{0.04}
\newcommand{\nxsres}{18.0}

\newcommand{\errth}{2\%}

\newcommand{\elepA}{182.7}
\newcommand{\smxsA}{0.26}
\newcommand{\xsA}{0.11}
\newcommand{\xsstA}{\ensuremath{\mbox{}^{0.16}_{0.11}}}
\newcommand{\xssyA}{0.04}

\newcommand{\elepB}{188.6}
\newcommand{\smxsB}{0.65}
\newcommand{\xsB}{0.69}
\newcommand{\xsstB}{\ensuremath{\mbox{}^{0.13}_{0.12}}}
\newcommand{\xssyB}{0.03}

\newcommand{\xsave}{0.67}
\newcommand{\xsavest}{0.13}
\newcommand{\xsavesy}{0.04}

\newcommand{\xsreB}{19.7}

\newcommand{\nctwo}{\ensuremath{{{\cal NC}\mathit{2}}}}
\newcommand{\nceight}{\ensuremath{{{\cal NC}\mathit{8}}}}

\newcommand{\wboson}{\ensuremath{\mathrm{W}}}

\newcommand{\nbkg}{\ensuremath{N_{\rm bkg}}}
\newcommand{\nobs}{\ensuremath{N_{\rm obs}}}

\newcommand{\chiw}{\ensuremath{\chi_{\mathrm{W}}}}
\newcommand{\chiz}{\ensuremath{\chi_{\mathrm{Z}}}}
\newcommand{\chii}{\ensuremath{\chi_i}}
\newcommand{\sigs}{\ensuremath{\sigma_S}}
\newcommand{\sigd}{\ensuremath{\sigma_D}}

\newcommand{\eff}{\ensuremath{\epsilon}}

\newcommand{\SM}{Standard Model\@}
\newcommand{\MC}{Monte Carlo\@}

\newcommand{\ALEPH}{{\sc aleph}}
\newcommand{\LEP}{{\sc lep}}
\newcommand{\CERN}{{\sc cern}}
\newcommand{\Aleph}{{ALEPH Collaboration}}

\newcommand{\jade}{{\sc jade}}
\newcommand{\durham}{{\sc durham}}

\newcommand{\etal}{{\it et al.}}

\newcommand{\pt}{\ensuremath{p_t}}

\newcommand{\yfszz}{{\sc yfszz}}
\newcommand{\pythia}{{\sc pythia}}
\newcommand{\excalibur}{{\sc excalibur}}
\newcommand{\grace}{{\sc grace4f}}
\newcommand{\koralw}{{\sc koralw}}
\newcommand{\koralz}{{\sc koralz}}
\newcommand{\phot}{{\sc phot02}}
\newcommand{\bhwide}{{\sc bhwide}}

\newcommand{\mz}{\ensuremath{m_{\zed}}}

\newcommand{\wlep}{\ensuremath{m_{\mathrm{lep}}}}
\newcommand{\whad}{\ensuremath{m_{\mathrm{had}}}}

\newcommand{\mmumu}{\ensuremath{m_{\mu\mu}}}
\newcommand{\mrec}{\ensuremath{m_{\rm rec}}}
\newcommand{\mll}{\ensuremath{m_{\ell\ell}}}
\newcommand{\mrecoil}{\ensuremath{m_{\rm recoil}}}

\newcommand{\mab}{\ensuremath{m_{12}}}
\newcommand{\mcd}{\ensuremath{m_{34}}}

\newcommand{\ycut}{\ensuremath{y_{\mathrm{cut}}}}
\newcommand{\ytf}{\ensuremath{y_{34}}}

\newcommand{\ecenn}{\ensuremath{f_{30^\circ}}}

\newcommand{\mvis}{\ensuremath{M_{\rm vis}}}

\newcommand{\ptvis}{\ensuremath{p_T}}
\newcommand{\pzvis}{\ensuremath{p_z}}
\newcommand{\emis}{\ensuremath{\not\!\!E}}
\newcommand{\mmis}{\ensuremath{\not\!\!M}}

\newcommand{\hacopm}{\ensuremath{A_{12}}}

\newcommand{\zed}{\ensuremath{{\mathrm{Z}}}}

\newcommand{\bq}{\ensuremath{{\mathrm{b}}}}

\newcommand{\nn}{\ensuremath{\nu\overline{\nu}}}
\newcommand{\qq}{\ensuremath{\mathrm{q}\overline{\mathrm{q}}}}

\newcommand{\bbbar}{\ensuremath{\mathrm{b}\overline{\mathrm{b}}}}

\newcommand{\ee}{\ensuremath{{\mathrm{e}^+\mathrm{e}^-}}}

\newcommand{\mm}{\ensuremath{\mu^+\mu^-}}
\newcommand{\tautau}{\ensuremath{\tau\tau}}
\newcommand{\tptm}{\ensuremath{\tau^+\tau^-}}

\newcommand{\qqqq}{\qq\qq}
\newcommand{\bbbb}{\bbbar\bbbar}
\newcommand{\bbqq}{\bbbar\qq}
\newcommand{\qqnn}{\qq\nn}
\newcommand{\bbnn}{\bbbar\nn}
\newcommand{\llnn}{\ensuremath{\ell\ell\nn}}
\newcommand{\llxx}{\ensuremath{\ell\ell\mathrm{XX}}}
\newcommand{\ttqq}{\ensuremath{\tautau\qq}}

\newcommand{\ww}{\ensuremath{\mathrm{WW}}}
\newcommand{\wen}{\ensuremath{\mathrm{We}\nu}}
\newcommand{\zz}{\ensuremath{\zed\zed}}
\newcommand{\zee}{\ensuremath{\zed\ee}}

\newcommand{\gaga}{\ensuremath{\gamma\gamma}}

\newcommand{\eisr}{\ensuremath{\gamma_\mathrm{peak}}}

\newcolumntype{d}[1]{D{.}{.}{#1}}

\begin{document}
 
\begin{titlepage}
\begin{center}
EUROPEAN LABORATORY FOR PARTICLE PHYSICS (CERN)
\end{center}
\begin{flushright}
CERN-EP/99-141\\
7 October 1999\\
\end{flushright}
\vfill
\begin{center}
  
\textbf{\boldmath\Large
Measurement of the $\ee\to\zz$ Production
Cross~Section at Centre-of-mass Energies of 183~and~189~\gev
\unboldmath
\\}

\mbox{}\\
ALEPH Collaboration\\

\mbox{}\\
\vfill

{\bf Abstract}
\end{center}

\noindent
The $\ee\to\zz$ cross section at $\roots=\elepA$ and \elepB~\gev\ 
has been measured using the \ALEPH\ detector.  The analysis covers
all of the visible \zz\ final states and yields cross section
measurements of
$$\sigma_{\zz}(\elepA~\gev) = \xsA \pm \xsstA\, \stat \pm \xssyA\,
\syst\, \pb$$ and $$\sigma_{\zz}(\elepB~\gev) = \xsave \pm \xsavest\,
\stat \pm \xsavesy\, \syst\, \pb,$$ consistent with the \SM\ expectations.

\vfill

\begin{center}
\textit{(To be submitted to Physics Letters B)}
\end{center}

\vfill
\end{titlepage}


\pagestyle{empty}
\newpage
\small
%
%
\newlength{\saveparskip}
\newlength{\savetextheight}
\newlength{\savetopmargin}
\newlength{\savetextwidth}
\newlength{\saveoddsidemargin}
\newlength{\savetopsep}
\setlength{\saveparskip}{\parskip}
\setlength{\savetextheight}{\textheight}
\setlength{\savetopmargin}{\topmargin}
\setlength{\savetextwidth}{\textwidth}
\setlength{\saveoddsidemargin}{\oddsidemargin}
\setlength{\savetopsep}{\topsep}
%
%
\setlength{\parskip}{0.0cm}
\setlength{\textheight}{25.0cm}
\setlength{\topmargin}{-1.5cm}
\setlength{\textwidth}{16 cm}
\setlength{\oddsidemargin}{-0.0cm}
\setlength{\topsep}{1mm}
\pretolerance=10000
\centerline{\large\bf The ALEPH Collaboration}
\footnotesize
\vspace{0.5cm}
{\raggedbottom
\begin{sloppypar}
\samepage\noindent
R.~Barate,
D.~Decamp,
P.~Ghez,
C.~Goy,
S.~Jezequel,
J.-P.~Lees,
F.~Martin,
E.~Merle,
\mbox{M.-N.~Minard},
B.~Pietrzyk
\nopagebreak
\begin{center}
\parbox{15.5cm}{\sl\samepage
Laboratoire de Physique des Particules (LAPP), IN$^{2}$P$^{3}$-CNRS,
F-74019 Annecy-le-Vieux Cedex, France}
\end{center}\end{sloppypar}
\vspace{2mm}
\begin{sloppypar}
\noindent
R.~Alemany,
S.~Bravo,
M.P.~Casado,
M.~Chmeissani,
J.M.~Crespo,
E.~Fernandez,
M.~Fernandez-Bosman,
Ll.~Garrido,$^{15}$
E.~Graug\`{e}s,
A.~Juste,
M.~Martinez,
G.~Merino,
R.~Miquel,
Ll.M.~Mir,
P.~Morawitz,
A.~Pacheco,
I.~Riu,
H.~Ruiz
\nopagebreak
\begin{center}
\parbox{15.5cm}{\sl\samepage
Institut de F\'{i}sica d'Altes Energies, Universitat Aut\`{o}noma
de Barcelona, 08193 Bellaterra (Barcelona), E-Spain$^{7}$}
\end{center}\end{sloppypar}
\vspace{2mm}
\begin{sloppypar}
\noindent
A.~Colaleo,
D.~Creanza,
M.~de~Palma,
G.~Iaselli,
G.~Maggi,
M.~Maggi,
S.~Nuzzo,
A.~Ranieri,
G.~Raso,
F.~Ruggieri,
G.~Selvaggi,
L.~Silvestris,
P.~Tempesta,
A.~Tricomi,$^{3}$
G.~Zito
\nopagebreak
\begin{center}
\parbox{15.5cm}{\sl\samepage
Dipartimento di Fisica, INFN Sezione di Bari, I-70126 Bari, Italy}
\end{center}\end{sloppypar}
\vspace{2mm}
\begin{sloppypar}
\noindent
X.~Huang,
J.~Lin,
Q. Ouyang,
T.~Wang,
Y.~Xie,
R.~Xu,
S.~Xue,
J.~Zhang,
L.~Zhang,
W.~Zhao
\nopagebreak
\begin{center}
\parbox{15.5cm}{\sl\samepage
Institute of High-Energy Physics, Academia Sinica, Beijing, The People's
Republic of China$^{8}$}
\end{center}\end{sloppypar}
\vspace{2mm}
\begin{sloppypar}
\noindent
D.~Abbaneo,
G.~Boix,$^{6}$
O.~Buchm\"uller,
M.~Cattaneo,
F.~Cerutti,
V.~Ciulli,
G.~Davies,
G.~Dissertori,
H.~Drevermann,
R.W.~Forty,
M.~Frank,
F.~Gianotti,
T.C.~Greening,
A.W.~Halley,
J.B.~Hansen,
J.~Harvey,
P.~Janot,
B.~Jost,
I.~Lehraus,
O.~Leroy,
P.~Maley,
P.~Mato,
A.~Minten,
A.~Moutoussi,
F.~Ranjard,
L.~Rolandi,
D.~Schlatter,
M.~Schmitt,$^{20}$
O.~Schneider,$^{2}$
P.~Spagnolo,
W.~Tejessy,
F.~Teubert,
E.~Tournefier,
A.E.~Wright
\nopagebreak
\begin{center}
\parbox{15.5cm}{\sl\samepage
European Laboratory for Particle Physics (CERN), CH-1211 Geneva 23,
Switzerland}
\end{center}\end{sloppypar}
\vspace{2mm}
\begin{sloppypar}
\noindent
Z.~Ajaltouni,
F.~Badaud
G.~Chazelle,
O.~Deschamps,
S.~Dessagne,
A.~Falvard,
C.~Ferdi,
\linebreak
P.~Gay,
C.~Guicheney,
P.~Henrard,
J.~Jousset,
B.~Michel,
S.~Monteil,
\mbox{J-C.~Montret},
D.~Pallin,
P.~Perret,
F.~Podlyski
\nopagebreak
\begin{center}
\parbox{15.5cm}{\sl\samepage
Laboratoire de Physique Corpusculaire, Universit\'e Blaise Pascal,
IN$^{2}$P$^{3}$-CNRS, Clermont-Ferrand, F-63177 Aubi\`{e}re, France}
\end{center}\end{sloppypar}
\vspace{2mm}
\begin{sloppypar}
\noindent
J.D.~Hansen,
J.R.~Hansen,
P.H.~Hansen,
B.S.~Nilsson,
B.~Rensch,
A.~W\"a\"an\"anen
\nopagebreak
\begin{center}
\parbox{15.5cm}{\sl\samepage
Niels Bohr Institute, 2100 Copenhagen, DK-Denmark$^{9}$}
\end{center}\end{sloppypar}
\vspace{2mm}
\begin{sloppypar}
\noindent
G.~Daskalakis,
A.~Kyriakis,
C.~Markou,
E.~Simopoulou,
A.~Vayaki
\nopagebreak
\begin{center}
\parbox{15.5cm}{\sl\samepage
Nuclear Research Center Demokritos (NRCD), GR-15310 Attiki, Greece}
\end{center}\end{sloppypar}
\vspace{2mm}
\begin{sloppypar}
\noindent
A.~Blondel,
\mbox{J.-C.~Brient},
F.~Machefert,
A.~Roug\'{e},
M.~Swynghedauw,
R.~Tanaka,
A.~Valassi,$^{23}$
\linebreak
H.~Videau
\nopagebreak
\begin{center}
\parbox{15.5cm}{\sl\samepage
Laboratoire de Physique Nucl\'eaire et des Hautes Energies, Ecole
Polytechnique, IN$^{2}$P$^{3}$-CNRS, \mbox{F-91128} Palaiseau Cedex, France}
\end{center}\end{sloppypar}
\vspace{2mm}
\begin{sloppypar}
\noindent
E.~Focardi,
G.~Parrini,
K.~Zachariadou
\nopagebreak
\begin{center}
\parbox{15.5cm}{\sl\samepage
Dipartimento di Fisica, Universit\`a di Firenze, INFN Sezione di Firenze,
I-50125 Firenze, Italy}
\end{center}\end{sloppypar}
\vspace{2mm}
\begin{sloppypar}
\noindent
M.~Corden,
C.~Georgiopoulos
\nopagebreak
\begin{center}
\parbox{15.5cm}{\sl\samepage
Supercomputer Computations Research Institute,
Florida State University,
Tallahassee, FL 32306-4052, USA $^{13,14}$}
\end{center}\end{sloppypar}
\vspace{2mm}
\begin{sloppypar}
\noindent
A.~Antonelli,
G.~Bencivenni,
G.~Bologna,$^{4}$
F.~Bossi,
P.~Campana,
G.~Capon,
V.~Chiarella,
P.~Laurelli,
G.~Mannocchi,$^{1,5}$
F.~Murtas,
G.P.~Murtas,
L.~Passalacqua,
M.~Pepe-Altarelli
\nopagebreak
\begin{center}
\parbox{15.5cm}{\sl\samepage
Laboratori Nazionali dell'INFN (LNF-INFN), I-00044 Frascati, Italy}
\end{center}\end{sloppypar}
\vspace{2mm}
\begin{sloppypar}
\noindent
M.~Chalmers,
L.~Curtis,
J.G.~Lynch,
P.~Negus,
V.~O'Shea,
B.~Raeven,
C.~Raine,
D.~Smith,
P.~Teixeira-Dias,
A.S.~Thompson,
J.J.~Ward
\nopagebreak
\begin{center}
\parbox{15.5cm}{\sl\samepage
Department of Physics and Astronomy, University of Glasgow, Glasgow G12
8QQ,United Kingdom$^{10}$}
\end{center}\end{sloppypar}
\begin{sloppypar}
\noindent
R.~Cavanaugh,
S.~Dhamotharan,
C.~Geweniger,$^{1}$
P.~Hanke,
V.~Hepp,
E.E.~Kluge,
A.~Putzer,
K.~Tittel,
S.~Werner,$^{19}$
M.~Wunsch$^{19}$
\nopagebreak
\begin{center}
\parbox{15.5cm}{\sl\samepage
Institut f\"ur Hochenergiephysik, Universit\"at Heidelberg, D-69120
Heidelberg, Germany$^{16}$}
\end{center}\end{sloppypar}
\vspace{2mm}
\begin{sloppypar}
\noindent
R.~Beuselinck,
D.M.~Binnie,
W.~Cameron,
P.J.~Dornan,$^{1}$
M.~Girone,
S.~Goodsir,
N.~Marinelli,
E.B.~Martin,
J.~Nash,
J.~Nowell,
H.~Przysiezniak,$^{1}$
A.~Sciab\`a,
J.K.~Sedgbeer,
E.~Thomson,
M.D.~Williams
\nopagebreak
\begin{center}
\parbox{15.5cm}{\sl\samepage
Department of Physics, Imperial College, London SW7 2BZ,
United Kingdom$^{10}$}
\end{center}\end{sloppypar}
\vspace{2mm}
\begin{sloppypar}
\noindent
V.M.~Ghete,
P.~Girtler,
E.~Kneringer,
D.~Kuhn,
G.~Rudolph
\nopagebreak
\begin{center}
\parbox{15.5cm}{\sl\samepage
Institut f\"ur Experimentalphysik, Universit\"at Innsbruck, A-6020
Innsbruck, Austria$^{18}$}
\end{center}\end{sloppypar}
\vspace{2mm}
\begin{sloppypar}
\noindent
C.K.~Bowdery,
P.G.~Buck,
G.~Ellis,
A.J.~Finch,
F.~Foster,
G.~Hughes,
R.W.L.~Jones,
N.A.~Robertson,
M.~Smizanska,
M.I.~Williams
\nopagebreak
\begin{center}
\parbox{15.5cm}{\sl\samepage
Department of Physics, University of Lancaster, Lancaster LA1 4YB,
United Kingdom$^{10}$}
\end{center}\end{sloppypar}
\vspace{2mm}
\begin{sloppypar}
\noindent
I.~Giehl,
F.~H\"olldorfer,
K.~Jakobs,
K.~Kleinknecht,
M.~Kr\"ocker,
A.-S.~M\"uller,
H.-A.~N\"urnberger,
G.~Quast,
B.~Renk,
E.~Rohne,
H.-G.~Sander,
S.~Schmeling,
H.~Wachsmuth
C.~Zeitnitz,
T.~Ziegler
\nopagebreak
\begin{center}
\parbox{15.5cm}{\sl\samepage
Institut f\"ur Physik, Universit\"at Mainz, D-55099 Mainz, Germany$^{16}$}
\end{center}\end{sloppypar}
\vspace{2mm}
\begin{sloppypar}
\noindent
J.J.~Aubert,
A.~Bonissent,
J.~Carr,
P.~Coyle,
A.~Ealet,
D.~Fouchez,
A.~Tilquin
\nopagebreak
\begin{center}
\parbox{15.5cm}{\sl\samepage
Centre de Physique des Particules, Facult\'e des Sciences de Luminy,
IN$^{2}$P$^{3}$-CNRS, F-13288 Marseille, France}
\end{center}\end{sloppypar}
\vspace{2mm}
\begin{sloppypar}
\noindent
M.~Aleppo,
M.~Antonelli,
S.~Gilardoni,
F.~Ragusa
\nopagebreak
\begin{center}
\parbox{15.5cm}{\sl\samepage
Dipartimento di Fisica, Universit\`a di Milano e INFN Sezione di
Milano, I-20133 Milano, Italy.}
\end{center}\end{sloppypar}
\vspace{2mm}
\begin{sloppypar}
\noindent
V.~B\"uscher,
H.~Dietl,
G.~Ganis,
K.~H\"uttmann,
G.~L\"utjens,
C.~Mannert,
W.~M\"anner,
\mbox{H.-G.~Moser},
S.~Schael,
R.~Settles,
H.~Seywerd,
H.~Stenzel,
W.~Wiedenmann,
G.~Wolf
\nopagebreak
\begin{center}
\parbox{15.5cm}{\sl\samepage
Max-Planck-Institut f\"ur Physik, Werner-Heisenberg-Institut,
D-80805 M\"unchen, Germany\footnotemark[16]}
\end{center}\end{sloppypar}
\vspace{2mm}
\begin{sloppypar}
\noindent
P.~Azzurri,
J.~Boucrot,
O.~Callot,
S.~Chen,
M.~Davier,
L.~Duflot,
\mbox{J.-F.~Grivaz},
Ph.~Heusse,
A.~Jacholkowska,$^{1}$
M.~Kado,
J.~Lefran\c{c}ois,
L.~Serin,
\mbox{J.-J.~Veillet},
I.~Videau,$^{1}$
J.-B.~de~Vivie~de~R\'egie,
D.~Zerwas
\nopagebreak
\begin{center}
\parbox{15.5cm}{\sl\samepage
Laboratoire de l'Acc\'el\'erateur Lin\'eaire, Universit\'e de Paris-Sud,
IN$^{2}$P$^{3}$-CNRS, F-91898 Orsay Cedex, France}
\end{center}\end{sloppypar}
\vspace{2mm}
\begin{sloppypar}
\noindent
G.~Bagliesi,
T.~Boccali,
C.~Bozzi,$^{12}$
G.~Calderini,
R.~Dell'Orso,
I.~Ferrante,
A.~Giassi,
A.~Gregorio,
F.~Ligabue,
P.S.~Marrocchesi,
A.~Messineo,
F.~Palla,
G.~Rizzo,
G.~Sanguinetti,
G.~Sguazzoni,
R.~Tenchini,
A.~Venturi,
P.G.~Verdini
\samepage
\begin{center}
\parbox{15.5cm}{\sl\samepage
Dipartimento di Fisica dell'Universit\`a, INFN Sezione di Pisa,
e Scuola Normale Superiore, I-56010 Pisa, Italy}
\end{center}\end{sloppypar}
\vspace{2mm}
\begin{sloppypar}
\noindent
G.A.~Blair,
J.~Coles,
G.~Cowan,
M.G.~Green,
D.E.~Hutchcroft,
L.T.~Jones,
T.~Medcalf,
J.A.~Strong
\nopagebreak
\begin{center}
\parbox{15.5cm}{\sl\samepage
Department of Physics, Royal Holloway \& Bedford New College,
University of London, Surrey TW20 OEX, United Kingdom$^{10}$}
\end{center}\end{sloppypar}
\vspace{2mm}
\begin{sloppypar}
\noindent
D.R.~Botterill,
R.W.~Clifft,
T.R.~Edgecock,
P.R.~Norton,
J.C.~Thompson,
I.R.~Tomalin
\nopagebreak
\begin{center}
\parbox{15.5cm}{\sl\samepage
Particle Physics Dept., Rutherford Appleton Laboratory,
Chilton, Didcot, Oxon OX11 OQX, United Kingdom$^{10}$}
\end{center}\end{sloppypar}
\vspace{2mm}
\begin{sloppypar}
\noindent
\mbox{B.~Bloch-Devaux},
P.~Colas,
B.~Fabbro,
G.~Fa\"if,
E.~Lan\c{c}on,
\mbox{M.-C.~Lemaire},
E.~Locci,
P.~Perez,
J.~Rander,
\mbox{J.-F.~Renardy},
A.~Rosowsky,
P.~Seager,$^{24}$
A.~Trabelsi,$^{21}$
B.~Tuchming,
B.~Vallage
\nopagebreak
\begin{center}
\parbox{15.5cm}{\sl\samepage
CEA, DAPNIA/Service de Physique des Particules,
CE-Saclay, F-91191 Gif-sur-Yvette Cedex, France$^{17}$}
\end{center}\end{sloppypar}
\vspace{2mm}
\begin{sloppypar}
\noindent
S.N.~Black,
J.H.~Dann,
C.~Loomis,
H.Y.~Kim,
N.~Konstantinidis,
A.M.~Litke,
M.A. McNeil,
G.~Taylor
\nopagebreak
\begin{center}
\parbox{15.5cm}{\sl\samepage
Institute for Particle Physics, University of California at
Santa Cruz, Santa Cruz, CA 95064, USA$^{22}$}
\end{center}\end{sloppypar}
\vspace{2mm}
\begin{sloppypar}
\noindent
C.N.~Booth,
S.~Cartwright,
F.~Combley,
P.N.~Hodgson,
M.~Lehto,
L.F.~Thompson
\nopagebreak
\begin{center}
\parbox{15.5cm}{\sl\samepage
Department of Physics, University of Sheffield, Sheffield S3 7RH,
United Kingdom$^{10}$}
\end{center}\end{sloppypar}
\vspace{2mm}
\begin{sloppypar}
\noindent
K.~Affholderbach,
A.~B\"ohrer,
S.~Brandt,
C.~Grupen,
J.~Hess,
A.~Misiejuk,
G.~Prange,
U.~Sieler
\nopagebreak
\begin{center}
\parbox{15.5cm}{\sl\samepage
Fachbereich Physik, Universit\"at Siegen, D-57068 Siegen, Germany$^{16}$}
\end{center}\end{sloppypar}
\vspace{2mm}
\begin{sloppypar}
\noindent
G.~Giannini,
B.~Gobbo
\nopagebreak
\begin{center}
\parbox{15.5cm}{\sl\samepage
Dipartimento di Fisica, Universit\`a di Trieste e INFN Sezione di Trieste,
I-34127 Trieste, Italy}
\end{center}\end{sloppypar}
\vspace{2mm}
\begin{sloppypar}
\noindent
J.~Putz,
J.~Rothberg,
S.~Wasserbaech,
R.W.~Williams
\nopagebreak
\begin{center}
\parbox{15.5cm}{\sl\samepage
Experimental Elementary Particle Physics, University of Washington, WA 98195
Seattle, U.S.A.}
\end{center}\end{sloppypar}
\vspace{2mm}
\begin{sloppypar}
\noindent
S.R.~Armstrong,
P.~Elmer,
D.P.S.~Ferguson,
Y.~Gao,
S.~Gonz\'{a}lez,
O.J.~Hayes,
H.~Hu,
S.~Jin,
J.~Kile,
P.A.~McNamara III,
J.~Nielsen,
W.~Orejudos,
Y.B.~Pan,
Y.~Saadi,
I.J.~Scott,
J.~Walsh,
\mbox{J.H.~von~Wimmersperg-Toeller},
Sau~Lan~Wu,
X.~Wu,
G.~Zobernig
\nopagebreak
\begin{center}
\parbox{15.5cm}{\sl\samepage
Department of Physics, University of Wisconsin, Madison, WI 53706,
USA$^{11}$}
\end{center}\end{sloppypar}
}
\footnotetext[1]{Also at CERN, 1211 Geneva 23, Switzerland.}
\footnotetext[2]{Now at Universit\'e de Lausanne, 1015 Lausanne, Switzerland.}
\footnotetext[3]{Also at Centro Siciliano di Fisica Nucleare e Struttura
della Materia, INFN Sezione di Catania, 95129 Catania, Italy.}
\footnotetext[4]{Also Istituto di Fisica Generale, Universit\`{a} di
Torino, 10125 Torino, Italy.}
\footnotetext[5]{Also Istituto di Cosmo-Geofisica del C.N.R., Torino,
Italy.}
\footnotetext[6]{Supported by the Commission of the European Communities,
contract ERBFMBICT982894.}
\footnotetext[7]{Supported by CICYT, Spain.}
\footnotetext[8]{Supported by the National Science Foundation of China.}
\footnotetext[9]{Supported by the Danish Natural Science Research Council.}
\footnotetext[10]{Supported by the UK Particle Physics and Astronomy Research
Council.}
\footnotetext[11]{Supported by the US Department of Energy, grant
DE-FG0295-ER40896.}
\footnotetext[12]{Now at INFN Sezione di Ferrara, 44100 Ferrara, Italy.}
\footnotetext[13]{Supported by the US Department of Energy, contract
DE-FG05-92ER40742.}
\footnotetext[14]{Supported by the US Department of Energy, contract
DE-FC05-85ER250000.}
\footnotetext[15]{Permanent address: Universitat de Barcelona, 08208 Barcelona,
Spain.}
\footnotetext[16]{Supported by the Bundesministerium f\"ur Bildung,
Wissenschaft, Forschung und Technologie, Germany.}
\footnotetext[17]{Supported by the Direction des Sciences de la
Mati\`ere, C.E.A.}
\footnotetext[18]{Supported by Fonds zur F\"orderung der wissenschaftlichen
Forschung, Austria.}
\footnotetext[19]{Now at SAP AG, 69185 Walldorf, Germany}
\footnotetext[20]{Now at Harvard University, Cambridge, MA 02138, U.S.A.}
\footnotetext[21]{Now at D\'epartement de Physique, Facult\'e des Sciences de Tunis, 1060 Le Belv\'ed\`ere, Tunisia.}
\footnotetext[22]{Supported by the US Department of Energy,
grant DE-FG03-92ER40689.}
\footnotetext[23]{Now at LAL, 91898 Orsay, France.}
\footnotetext[24]{Supported by the Commission of the European Communities,
contract ERBFMBICT982874.}
%
%
\setlength{\parskip}{\saveparskip}
\setlength{\textheight}{\savetextheight}
\setlength{\topmargin}{\savetopmargin}
\setlength{\textwidth}{\savetextwidth}
\setlength{\oddsidemargin}{\saveoddsidemargin}
\setlength{\topsep}{\savetopsep}
\normalsize
\newpage
\pagestyle{plain}
\setcounter{page}{1}


\setcounter{page}{1}
 
\section{Introduction}

\suppressfloats

\begin{figure}[t]
\begin{center}
\begin{tabular}{cc}
\epsfbox{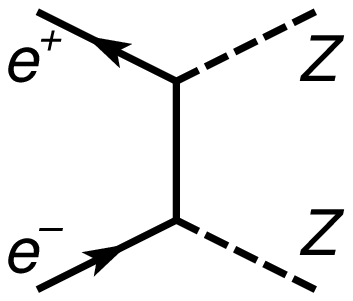} & \epsfbox{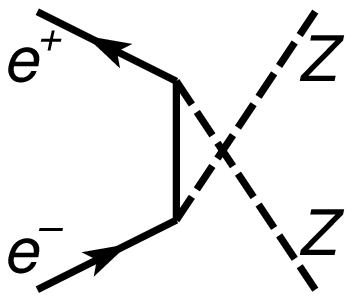} \\
\end{tabular}
\end{center}
\caption{The 
$t$-channel exchange diagrams show the $\ee\to\zz$ \nctwo\
processes in the \SM.  
\label{fig:feyn}}
\end{figure}

The successful operation of \LEP\ at and above the \zz\ threshold in
1997 and 1998 allows observation of a sizable number of pair-produced
resonant \zed\ bosons.  Within the \SM, the process $\ee\to\zz$
proceeds via the two \nctwo\ diagrams which involve the $t$-channel
exchange of an electron (see Figure~\ref{fig:feyn}).  A measurement of
this process, the dominant irreducible background for Higgs boson
searches at \LEP, tests the rates predicted by the \SM\@.

This letter describes a measurement of the \zz\ \nctwo\ cross section
at a centre-of-mass energy of \elepB~\gev\ using the \ALEPH\ detector.
It begins with a description of the detector and data samples,
continues with a description of the event selections for the various
final states, and concludes with the measured cross section.
Application of the same analysis to the 1997 \ALEPH\ data also yields
a measurement of the cross section at $\roots=$\elepA~\gev.


\section{ALEPH Detector}

This section briefly describes the most salient features of the
\ALEPH\ detector; detailed descriptions of the detector and its
performance can be found elsewhere~\cite{detectorA,detectorB}.

\ALEPH\ is a cylindrically symmetric detector with its axis coincident
with the beam axis.  The three innermost detectors---a silicon
microstrip detector, the inner tracking chamber, and the
time-projection chamber (TPC)---measure the momentum of charged
particles.  With a 1.5~T axial magnetic field provided by a
superconducting solenoidal coil, these detectors together achieve a
transverse momentum resolution of $\delta \pt /\pt = 6 \times
10^{-4}\,\pt \oplus 5 \times 10^{-3}$ (\pt\ in \gevc).

The excellent resolution of the silicon microstrip detector
facilitates the identification of jets containing \bq~hadrons. A
neural network combines impact parameter and secondary vertex
information with other jet information such as track multiplicity,
track rapidity, and the presence of leptons, to provide a measure of
the \bq~content of jets.

An electromagnetic calorimeter placed between the TPC and the
superconducting coil identifies electrons and photons, and measures
their energies with a resolution of $\delta E/E = 0.18/\sqrt{E}+0.009$
($E$ in \gev). This sampling calorimeter has a depth of 22 radiation
lengths and consists of projective towers each of which subtends a
solid angle of approximately $0.9^\circ \times 0.9^\circ$.

The iron return yoke is instrumented with 23 layers of streamer tubes
and serves as a hadronic calorimeter.  It is seven interaction lengths
deep and achieves an energy resolution of $\delta E/E = 0.85/\sqrt{E}$
for charged and neutral hadrons.  Two additional layers of streamer
tubes outside the return yoke aid the identification of muons.


The luminosity is measured with small-angle Bhabha events, using
lead-proportional wire sampling calorimeters~\cite{detectorC}, with an
accepted Bhabha cross section of 4.6~\nb\ at \elepA~\gev\ and 4.3~\nb\
at \elepB~\gev~\cite{BHLUMI}.  Integrated luminosities of $56.81\pm
0.11\,\stat \pm 0.29\,\syst$~\invpb\ and $174.21 \pm 0.20\,\stat \pm
0.82\,\syst$~\invpb\ were recorded respectively, at mean
centre-of-mass energies of $182.66 \pm 0.05$~\gev\ and $188.63 \pm
0.04$~\gev~\cite{LEPbeam}.

An energy flow algorithm~\cite{detectorB}\ combines the information
from the tracking detectors and calorimeters and provides a list of
reconstructed charged and neutral particles.  These energy flow
objects are used in the analysis described below.

\section{Simulation of Signal and Background}

The \yfszz\ \MC\ generator~\cite{yfszz}\ provides a calculation of the
expected \SM\ cross section for the \nctwo\ processes.  The expected
cross section is \smxsA~\pb\ at $\roots=\elepA$~\gev\ and \smxsB~\pb\
at \elepB~\gev.  With a consistent set of electroweak parameters,
three other \MC\ generators---\pythia~\cite{pythia},
\excalibur~\cite{excalibur}, and \grace~\cite{grace}---all agree with
the \yfszz\ calculation within \errth. This value is taken as the
theoretical uncertainty on the \nctwo\ cross section determination.

Samples of \zz\ compatible events from the \nctwo\ and \nceight\
diagrams were generated with the \pythia\ \MC\ generator.  The
efficiencies were determined from the \nctwo\ sample.  The \nceight\
diagrams include, in addition to the diagrams of
Figure~\ref{fig:feyn}, six other diagrams in which one or both \zed\
bosons are replaced by a virtual photon.  The difference between the
\nctwo\ sample and the \nceight\ sample was taken as background.  This
procedure accounts for any interference between the four-fermion
diagrams.  As a check, the analysis was repeated with samples
generated with the \excalibur\ \MC\ generator (see
Section~\ref{xcheck:a}).

The \koralw~\cite{koralw}\ \MC\ generator was used to create a sample
of \ww-like four-fermion background events.  To avoid double counting,
events in the \nceight\ \zz\ sample with a \ww-like final state are
removed.  As the \koralw\ generator does not include the part of the
$\ee\to\wen$ and $\ee\to\zee$ processes with an electron near the beam
axis and outside of the detector fiducial volume, additional samples
for these processes were generated with the \pythia\ \MC\ generator.

For the two-fermion background, $\ee\to\mm$ and $\ee\to\tptm$ events
were generated with \koralz~\cite{koralz}, $\ee\to\ee$ events with
\bhwide~\cite{bhwide}, and $\ee\to\qq$ events with \pythia\@.  The
two-photon background---$\gaga\to\ee$, $\gaga\to\mm$, and
$\gaga\to\tptm$---was generated with \phot~\cite{phot}.  Studies
showed that the process $\gaga\to {\rm hadrons}$ contributes
negligibly to the background expectation after all of the analysis
cuts.

\section{Event Selection}

The analyses described below select events from all visible \zz\ final
states. While based on the Higgs boson search~\cite{higgs}, these
analyses have been tailored to the \zz\ production process.

Two parallel analyses measure the \zz\ production cross section.  The
first is an entirely cut-based analysis which uses the \qq\qq, \qq\nn,
\llxx, and \llnn\ channels.  Throughout this paper, the symbol
$\ell$ denotes an electron or muon and X denotes a quark or charged
lepton.  The second analysis, the ``NN'' analysis, replaces the
cut-based selections in the \qq\qq\ and \qq\nn\ channels with neutral
networks and introduces an additional neural network to identify the
$\ttqq$ final state.

Table~\ref{tab:summary}\ summarizes the results of the selections
described below.

\begin{table}
\begin{center}
\caption{The efficiencies, \zz\ signal expectation, background
expectation, number of observed events, and the measured \nctwo\ cross
section with the statistical error for each channel at $\roots=
\elepB$~\gev\@.\label{tab:summary}}
\begin{tabular}{ld{1}@{$\pm$}d{1}d{1}d{1}@{$\pm$}d{1}rd{2}@{$\pm$}l}
\multicolumn{9}{c}{\mbox{}}\\
\hline\hline
& \multicolumn{2}{c}{\eff\ (\%)}
& \multicolumn{1}{c}{$N^{\rm SM}_{\zz}$}
& \multicolumn{2}{c}{\nbkg}
& \multicolumn{1}{c}{\nobs} 
& \multicolumn{2}{c}{$\sigma_{\zz}$~(\pb)} \\
\hline
\llxx$\mbox{}^\ddagger$     &  76.7 & 0.5 & 8.8 &  1.1 & 0.3  & 12  &  0.80 & $\mbox{}^{0.28}_{0.23}$ \\
\qqqq\ (\bq)$\mbox{}^*$     &  14.6 & 0.2 & 8.3 &  3.8 & 0.4  & 14  &  0.80 & $\mbox{}^{0.32}_{0.27}$ \\
\qqqq\ (non-\bq)$\mbox{}^*$ &  27.4 & 0.2 &16.0 & 60.0 & 1.4  & 69  &  0.36 & $\mbox{}^{0.35}_{0.32}$ \\
\qqqq$\mbox{}^\dagger$      &  31.5 & 1.2 &17.2 & 19.7 & 1.8  & 32  &  0.58 & $\mbox{}^{0.22}_{0.19}$ \\
\qqnn$\mbox{}^*$            &  47.2 & 0.3 &15.0 & 13.0 & 0.6  & 30  &  0.74 & $\mbox{}^{0.25}_{0.22}$ \\
\qqnn$\mbox{}^\dagger$      &  79.8 & 0.7 &25.3 & 64.5 & 1.6  & 88  &  0.65 & $\mbox{}^{0.22}_{0.20}$ \\
\llnn$\mbox{}^\ddagger$     &  50.3 & 1.5 & 1.4 &  1.5 & 0.3  &  1  & -0.26 & $\mbox{}^{0.66}_{0.30}$ \\
\ttqq$\mbox{}^\dagger$      &  42.2 & 1.3 & 2.2 &  1.5 & 0.2  &  3  &  0.44 & $\mbox{}^{0.62}_{0.41}$ \\
\hline\hline 
\multicolumn{9}{l}{$\mbox{}^*\,$cut analysis
\qquad$\mbox{}^\dagger\,$NN analysis
\qquad$\mbox{}^\ddagger\,$both analyses} \\
\end{tabular}
\end{center}
\end{table}

\subsection{\boldmath Selection of $\zz\to\llxx$ Final States\unboldmath}

This analysis identifies those events in which one of the \zed\ bosons
decays into a pair of electrons or muons and the other decays into
anything except neutrinos.  Although the branching fraction of
$\zed\to\ell\ell$ is small, the analysis benefits greatly from the
high lepton identification efficiency and excellent mass resolution.

Selected events have four or more reconstructed charged particles.
The total energy of all the charged particles must exceed 10\% of the
centre-of-mass energy.  To remove radiative returns to the \zed, the
most energetic isolated photon in an event must have an energy less
than $0.75\,\eisr$, where \eisr\ is the most likely energy of the
radiative return photon and is given by $\eisr = 0.5\,(\roots -
m_\zed^2/\roots)$.

Standard lepton cuts are used to identify electrons and muons in the
event~\cite{leptons}.  Electron energies are corrected for possible
bremsstrahlung losses.  To maintain a high selection efficiency,
isolated charged particles are also considered as lepton candidates.
The isolation is defined as the half-angle of the largest cone about a
particle's direction which contains 5\% or less of the event's total
energy.  Isolated particles have an isolation angle greater than
$10^\circ$.  The pair of oppositely-charged lepton candidates with an
invariant mass closest to the \zed\ mass is chosen as the
$\zed\to\ell\ell$ pair.  Only those pairs with at least one identified
lepton are considered and $e\mu$ combinations are rejected. Photons
consistent with final state radiation are included in the calculation
of the invariant mass of the lepton pair.  Events in which the lepton
pair contains an electron consistent with one from a photon conversion
are rejected.

After the $\zed\to\ell\ell$ lepton pair has been selected, the
\durham\ algorithm~\cite{durham}\ clusters the remainder of the event
into two jets; these jets must contain at least one charged particle.
To remove $\zed\gamma^\ast$ events, the selection requires that the
invariant mass of the two jets exceed 15~\gevcc.  To remove \qq\
events, the sum of the transverse momenta of the leptons with respect
to the nearest jet must be greater than 20~\gevc, where the nearest
jet is the one which forms the smallest invariant mass with the
lepton.

The process $\ww\to\qq\,\ell\nu$ constitutes a large background for
the selected events in which the $\zed\to\ell\ell$ pair contains one
identified lepton and one isolated charged particle. The two \wboson\
masses are calculated assuming that the event comes from this process.
The mass \wlep\ is the invariant mass of the identified lepton and the
missing momentum; the other mass \whad\ is the invariant mass of the
remainder of the event.  For selected events, these masses satisfy
either $\whad+\wlep<150$~\gevcc\ or $\whad-\wlep>20$~\gevcc.

\begin{figure}[t]
\begin{center}
\epsfxsize=0.8\textwidth\epsfbox{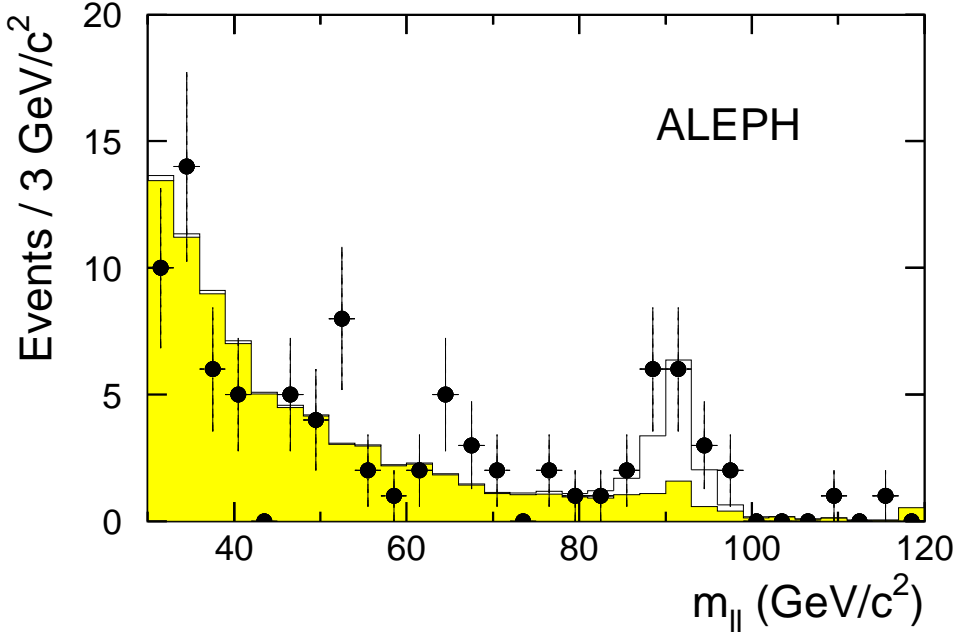}
\end{center}
\caption{The distribution of the dilepton masses for the \llxx\
channel.  All cuts except the elliptical mass cut have been applied.
The grey histogram shows the expected background and the hollow
histogram shows the expected signal.  \label{fig:massllxx}}
\end{figure}

The mass of the lepton pair is used as the mass of the first \zed\ 
candidate; that of the second candidate is the mass recoiling against
the two leptons. Requiring the above cuts and that both \zed\ masses
be larger than 30~\gevcc, the analysis selects 92 events from the data
with $90.6\pm 1.4$ expected from signal plus background.
Figure~\ref{fig:massllxx}\ shows the dilepton invariant mass
distribution.

\begin{figure}[t]
\begin{center}
\epsfxsize=\hsize\epsfbox{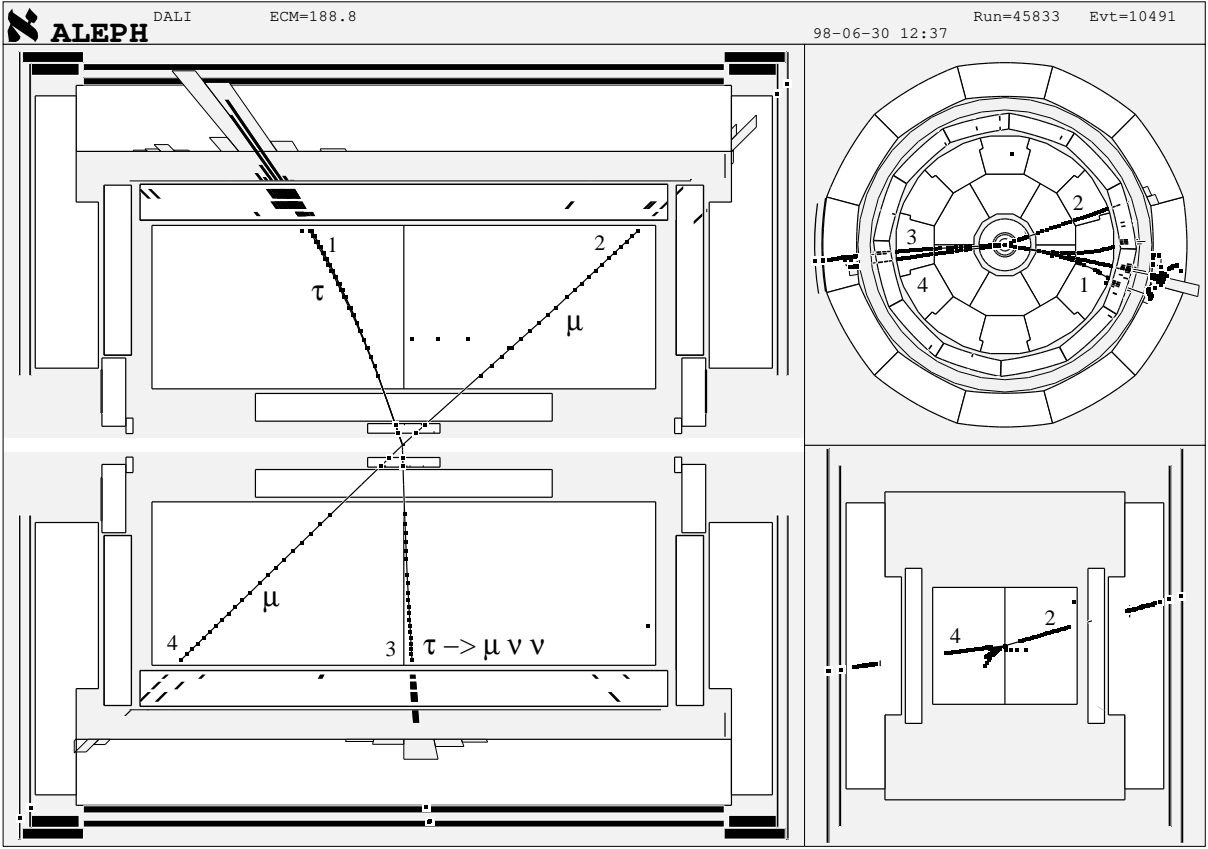}
\end{center}
\caption{An event selected by the \llxx\ analysis and classified as a
$\zz\to\mu\mu\, \tautau$ candidate.  Tracks 2 and 4 form the
$\zed\to\mu\mu$ candidate with an invariant mass of 96.0~\gevcc.  Tracks
1 and 3 form the $\zed\to\tautau$ candidate with a recoil mass of
90.3~\gevcc.\label{zzevt}}
\end{figure}

To further reduce background levels, the two reconstructed masses must
be consistent with the \zed\ boson mass. As the leptonic invariant
mass and recoil mass have different resolutions, an elliptical cut is
defined using
$$
 r^2 = \left(\frac{\mll - \mz}{\sigma_{\mll}}\right)^2 +
\left(\frac{\mrecoil - \mz}{\sigma_{\mrecoil}}\right)^2\,,
$$
where $\sigma_{\mll}=2.5$~\gevcc\ and $\sigma_{\mrecoil}=3.3$~\gevcc.
Selected events have $r<3$.  An \llxx\ candidate, classified as a
$\zz\to\mu\mu\, \tautau$ event, is shown in Figure~\ref{zzevt}.

Systematic uncertainties from lepton identification, tracking
resolutions, and reconstruction of kinematic variables have been
studied.  The total relative systematic uncertainty on the efficiency
is 1.3\% and the relative uncertainty on the expected number of
background events is 27\%.  Both uncertainties are dominated by
limited \MC\ statistics.

\subsection{\boldmath Selection of $\zz\to\qq\qq$ Final States\unboldmath}

A total reconstructed energy near \roots\ and four distinct jets
characterize events from the $\zz \to \qqqq$ final state. This final
state has the largest branching fraction, but also the largest
background.

The cut-based and NN-based analyses of this channel use a common
preselection.  The event must contain eight or more charged particles
and the total energy of all charged particles must exceed 10\% of the
centre-of-mass energy.

The \durham\ algorithm clusters the event into four jets.  To remove
those events inconsistent with a four-jet topology, the value of \ytf,
the \ycut\ in the clustering algorithm which changes the event from a
three-jet to four-jet topology, must be larger than 0.004.
Additionally, each jet must contain at least one charged track.

Two additional cuts suppress events with initial state radiation.
First, the longitudinal momentum of the event must satisfy the
requirement $\left|\pzvis\right| < 1.5\,(\mvis-90)$ with \mvis\ in
\gevcc\ and \pzvis\ in \gevc, to remove events in which the initial
state photon is along the beam axis and outside of the detector
fiducial volume.  Second, the electromagnetic fraction of all of the
jets must be smaller than 80\% to remove those events in which the
photon is inside the fiducial volume. This second cut also removes the
\qq\ee\ overlap with the leptonic channel described above.

\subsubsection{\boldmath Cut-based Selection\unboldmath}

The cut analysis tightens the preselection with three additional cuts.
The first two target \qq\ events: the thrust value of the event must
be smaller than 0.9 and the sum of the four smallest angles between
jets must be greater than $350^\circ$.  The third cut reduces the
overlap of \qq\mm\ events between the \qq\qq\ and the \llxx\
selections. Accepted events have an invariant mass of the two most
energetic muon candidates \mmumu\ less than 50~\gevcc\ and satisfy the
condition $p_1+p_2-\mmumu c <35$~\gevc, where $p_1$ and $p_2$ are the
momenta of the two muons.  For events with only one identified muon,
this cut reduces to $p_1<35$~\gevc.  These negligibly affect the
efficiency for the \qq\qq\ final state but reject more than 97\% of
the \qq\mm\ overlap and also a small fraction of the \ww\ background.

After the above preselection, $1352\pm 6$ events are expected from the
simulation, approximately 80\% of which are $\ww \to \qqqq$ events. In
the data, 1219 events are observed. This discrepancy has already been
reported by \ALEPH\ in a preliminary measurement of the $\ww \to
\qqqq$ cross section~\cite{whadr}.  Its impact on the \zz\ cross
section measurement is studied in Section~\ref{xcheck:b}.

The selection subsequently relies on the tagging of jets from
\bq~quarks and on the dijet mass information.  A four-constraint
fit~\cite{wmass} is then applied to improve the jet momentum
resolution.  A \bq-tagging neural network produces a value $\eta_i$
for each jet which is near unity for \bq~jets and near zero for other
jets~\cite{higgs}.

The four-jet analysis is split into three sub-analyses: a \bbbb\ 
selection, a \bbqq\ selection, and a non-\bq~quark selection.  When
calculating the cross section, the two \bq~selections are treated as a
single channel.  The last sub-analysis replaces the \bq-tagging cuts
with strict mass requirements to reject background while retaining
efficiency for the majority of \zz\ events ($\sim$62\%) which do not
contain \bq~quark jets.

The \bbbb\ analysis selects events with high \bq~content,
well-isolated jets, and large dijet masses. It requires that $y_{34} >
0.020$, that the sum of the dijet masses for at least one of the dijet
combinations be above 170~\gevcc, and that $9.5\,y_{34}+\sum\eta_i >
3.1$ where the sum is over the four reconstructed jets.

For the other sub-selections the \bq-tagging requirements give way to
tighter dijet mass requirements.  The mass information is contained in
the quantities \chiw\ and \chiz\ defined as
$$\chii^2=\left(\frac{\mab+\mcd-2\,m_i}{\sigs^i}\right)^2
+\left(\frac{\mab-\mcd}{\sigd^i}\right)^2\,,$$ where $i$ stands for
\wboson\ or \zed\@. The quantities $\sigs^i$ and $\sigd^i$ are the
resolutions of the sum and the difference of the dijet masses for the
correct dijet combination, respectively.  The contours of constant
\chiw\ and \chiz\ define two ellipses referred to as \ww\ and \zz\
ellipses throughout this section. For the \ww\ and \zz\ ellipses
respectively, the values 4 and 3~\gevcc\ are used for $\sigs^i$ and
the values 10 and 16~\gevcc, for $\sigd^i$.

The \bbqq\ selection requires that at least one dijet combination
falls inside the \zz\ ellipse with $\chiz < 2.40$. For that
combination, the dijet not containing the most poorly \bq-tagged jet
must be compatible with $\zed\to\bbbar$ in terms of \bq-tagging:
$\min(\eta_1,\eta_2)>0.20$ and $-\log_{10}(1-\eta_1)(1-\eta_2) >
1.50$.

Finally, the selection for \qqqq\ events without \bq~jets raises the
\ytf\ cut to 0.006, tightens further the \zz\ ellipse to $\chiz <
1.75$, and reduces the \ww\ background by requiring that no dijet
combination falls inside a \ww\ ellipse with $\chiw < 1.60$.  To
maintain the statistical independence of the \bq\ and non-\bq\ 
selections, this selection accepts only those events which have not
been previously selected by the \bbbb\ or \bbqq\ cuts.

\begin{figure}[t]
\begin{center}
\epsfxsize=0.8\textwidth\epsfbox{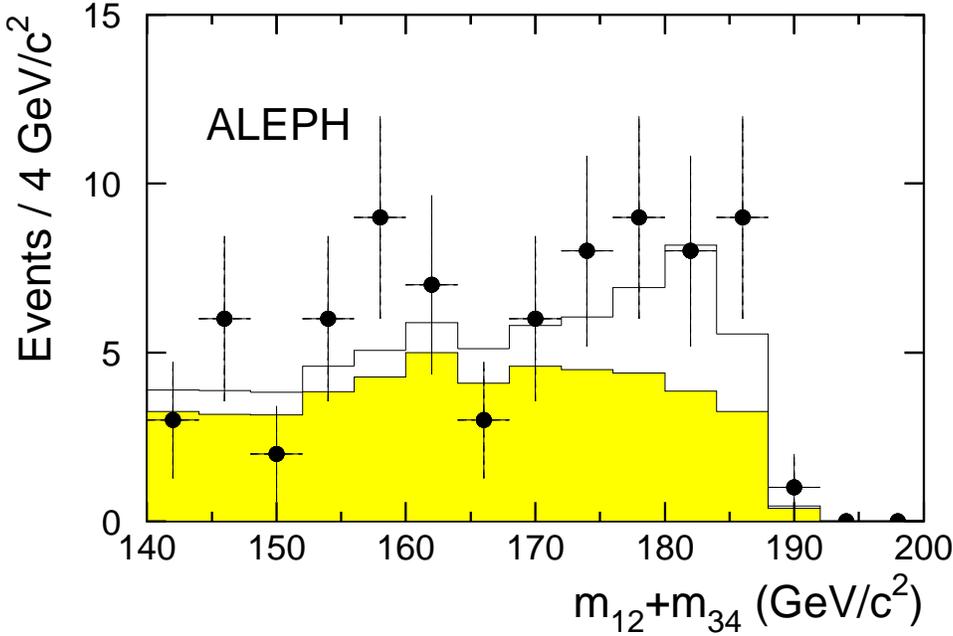}
\end{center}
\caption{The distribution of the sum of the dijet masses
  ($m_{12}+m_{34}$) for the \bbbb\ and \bbqq\ channels.  All
  cuts have been applied except the elliptical mass cut. The hollow
  histogram is the contribution from the signal and the darker region
  that from the expected background.\label{fig:massbbqq}}
\end{figure}

Figure~\ref{fig:massbbqq}\ shows the sum of the dijet masses for the
\bbbb\ and \bbqq\ selections with the elliptical mass cut removed.

Although this selection does not specifically target \tautau\qq\ 
events, it nonetheless has a significant efficiency for these events
($\sim$17\%). Consequently, the \tautau\qq\ selection described below
is not included in the cut-based cross section measurement.

\subsubsection{Neural Network Selection}

\begin{figure}[t]
\begin{center} 
\epsfxsize=0.8\textwidth\epsfbox{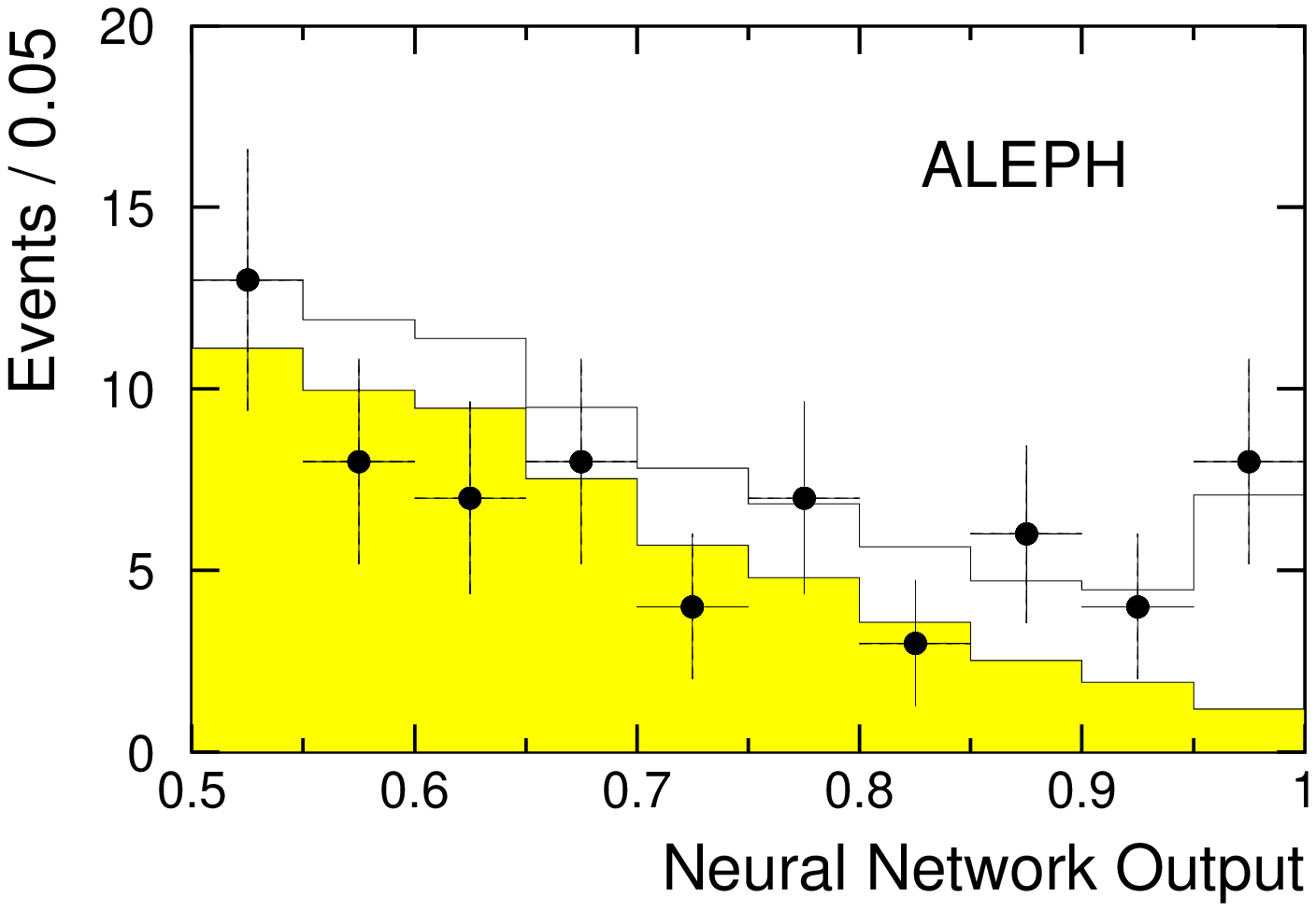}
\end{center}
  \caption{Neural network output from the \qqqq\ selection for data
    and Monte Carlo simulation.  The grey histogram shows the expected
    background, and the hollow histogram shows the expected \nctwo\ 
    signal.\label{fig:jetNN}}
\end{figure}

A multivariate neural network is trained to select \zz\ signal events
from \qq\ and \ww\ background events.  During training, the network is
presented with up to six dijet combinations per \qq\ or \ww\ event,
but only the correct dijet pairing for each signal event.  If one
\zed\ boson decays to \bq~quarks, those jets are labelled 3 and 4, and
the other two jets are labelled 1 and 2.  Otherwise the jet labels are
random.  Reconstructed dijet masses are obtained using a
four-constraint fit which requires energy-momentum conservation.
Twenty-three event variables are used to discriminate specific signal
and background features.  In many cases, inclusion of similar
variables in different combinations improves the neural network
training.

The event variables \ytf, thrust, and sphericity aid in selecting
events with four separated jets.  The missing energy and
$\max\left({E_{\mathrm{charged\ track}}/E_{\mathrm{jet}}} \right)$,
the maximum scaled energy among all jets of the most energetic charged
track in a jet, specifically reject semileptonic \wboson\ decays.

A \bq-tagging neural network~\cite{higgs} gives an output $\eta_i$ for
each jet, and these variables are input as $\min(\eta_3,\eta_4)$,
$(1-\eta_3)(1-\eta_4)$, and $\sum^{4}_{i=1}\eta_i$.

The reconstructed dijet masses aid the selection of resonant \zz\ 
production events.  The mass combinations used as network inputs are
$m_{12}$, $m_{34}$, and $(m_{12}-m_{34})^2 + (m_{34}-m_{\zed})^2$.
These variables perform the same function as an elliptical mass cut.

The jet boosted sphericity (calculated in the rest frame of the jet)
of the two jets with lowest $\eta_i$, the track multiplicity in these
jets, and the two lowest jet masses all help to separate light quark
jets from gluon jets.  The track multiplicity in this case is
restricted to tracks with a rapidity with respect to the jet axis
greater than 1.6.

The sum of the four smallest inter-jet angles discriminates against
\qq\ background events.  The angular variable $\min \left( \cos
  \theta_{ij} + \cos \theta_{kl} \right)$ minimized over all jet
combinations selects events with pairs of back-to-back jets.  The
event broadening variable $B=\min(B_+, B_-)$ also aids in this
discrimination by rejecting \qq\ background with gluons that can lead
to broad jets.  The broadening variable for each of the two
hemispheres $S_\pm$ is given by $B_\pm = (\sum_{i\,\in\,S_\pm}
  p_{Ti})/(\sum_i \left| p_i \right|)$, where \ptvis\ 
is calculated with respect to the thrust axis and the sums run over
tracks.  Finally, the largest jet energy and the two smallest jet
energies improve the overall discriminating power.

Of all of the dijet combinations, the one with the largest neural
network output is used. The output of the neural network is shown in
Figure~\ref{fig:jetNN} for Monte Carlo simulation and for data.  Only
events with a network output greater than 0.7 are used in the cross
section calculation.

\subsubsection{Systematic Uncertainties}

The systematic uncertainties in this channel include effects from the
modelling of \bq~physics, from discrepancies in tracking between the
simulation and the data, from discrepancies in reconstructed jet
kinematics, from uncertainties in the background cross sections, and
from uncertainties in gluon splitting into heavy flavours.  

For the cut-based analysis, the total relative systematic
uncertainties on the signal efficiencies are 1.7\% and 1.3\% for the
\bq\ and non-\bq\ selections, respectively, with all of the sources
having comparable contributions.  The relative uncertainties for the
background are 13\% and 4\%, dominated by the limited \MC\ statistics
and jet corrections.

For the NN-based analysis, the total relative systematic uncertainty
on the signal efficiency is 3\%.  No single effect dominates the
uncertainty.  The total relative uncertainties for the background are
11\% (\qq) and 7\% (\ww), dominated by uncertainties in the
reconstruction of kinematic variables and limited Monte Carlo
statistics.

\subsection{\boldmath Selection of $\zz\to\qq\nn$ Final States\unboldmath}

Approximately 30\% of the \zz\ events have a \qq\nn\ final state
characterized by a missing mass \mmis\ and visible mass consistent
with the \zed\ mass.

Both the cut-based and NN-based analyses for this channel share a
common preselection.  Events must contain more than four reconstructed
charged particles and the total energy of all the charged particles
must exceed 10\% of the centre-of-mass energy.  The plane
perpendicular to the thrust axis divides the event into two
hemispheres.  These hemispheres are the ``jets'' used to calculate all
of the kinematic quantities below.  Both of the hemispheres must have
a nonzero energy.

The preselection rejects events with a total energy more than
$30^\circ$ degrees from the beam axis smaller than $0.25\,\roots$ and
a missing transverse momentum smaller than $0.05\,\roots$.  This cut
primarily removes $\gaga\to \mathrm{hadrons}$ events.

Two additional cuts remove much of the background from \qq\ events
with initial state radiation.  First, the magnitude of the
longitudinal event momentum must be less than 50~\gevc.  Secondly, the
missing mass must be greater than 50~\gevcc.

After the preselection, the background is dominated by \qq\ and \ww\
events.  Assuming the \SM\ cross section for \zz\ production, $1533\pm
7$ events are expected in good agreement with the 1493 actually observed
in the data. 

\subsubsection{\boldmath Cut-based Selection\unboldmath}

The two-fermion background remaining after the preselection consists
largely of \qq\ events with two or more initial state radiative
photons.  Requiring the acoplanarity \hacopm\ to be greater than 0.08
removes much of this background.  The acoplanarity of the two jets is
$\hacopm=(\hat{\jmath}_1 \times \hat{\jmath}_2) \cdot \hat{z}$ where
$\hat{\jmath}_i$ are the unit vectors along the jet directions.

\begin{figure}[t]
\begin{center}
\epsfxsize=0.8\textwidth\epsfbox{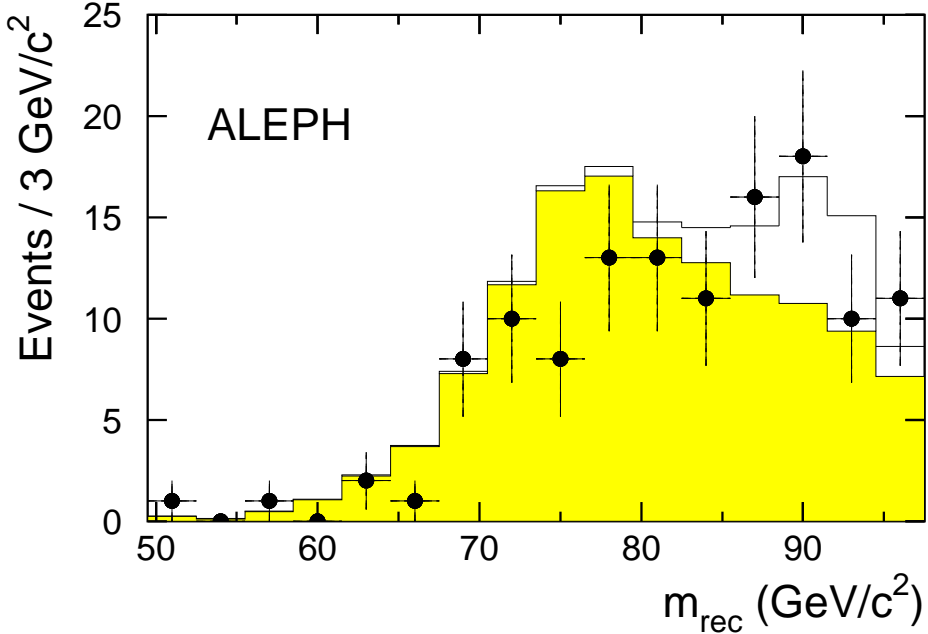}
\end{center}
\caption{The distribution of reconstructed masses for the \qqnn\
  channel. The grey region is the expected contribution from the
  background and the hollow region that from the \zz\ \nctwo\ 
  signal.\label{fig:massqqnn}}
\end{figure}

Pairs of \wboson\ bosons which subsequently decay as $\ww \to
\tau\nu\,\qq$ form the other major background.  To reject cases where
the tau lepton decays leptonically, a cut on the isolation of
identified leptons is made.  The isolation is defined as the sum of
the energy in a $30^\circ$ cone around the lepton direction.  Selected
events have an isolation greater than 13~\gev.  When the lepton from
the tau decay is unidentified or the tau lepton decays hadronically,
the previous cut is ineffective.  To remove these cases, the analysis
requires that the angle between the reconstructed charged particle
with the highest momentum and the nearest charged particle be less
than $20^\circ$.  Additionally, the event is reclustered into minijets
with the \jade\ algorithm~\cite{jade}\ using
$\ycut=(2~\gev/\roots)^2$.  The energy of the most isolated minijet
must be smaller than 8~\gev.

To remove events from the $\ee\to\wen$ and $\ee\to\zee$ processes
which have a detected final state electron near the beam axis, the
analysis requires that the energy in a cone of $12^\circ$ around the
beam axis must be smaller than $0.02\,\roots$.  To remove events which
have a missing momentum near the beam axis and may be poorly measured,
the smallest angle between the missing momentum and the beam axis
must be larger than $25^\circ$.  Figure~\ref{fig:massqqnn}\ 
shows the distribution of the reconstructed mass for candidates which
pass all of the above cuts.  The reconstructed mass \mrec\ is the
invariant mass of the two jets calculated with the missing mass
constrained to \mz.

Because the mass of the \zed\ boson is known, the mass information can
be used effectively to reject much of the background.  As for the
other channels, this selection uses an elliptical mass cut of
``radius''
$$r^2 = \left(\frac{\mrec-\mz}{\sigma_{\mrec}}\right)^2 +
\left(\frac{\mmis-\mz}{\sigma_{\not M}}\right)^2\,,$$ where
$\sigma_{\mrec}=3.1$~\gevcc\ and $\sigma_{\mmis} = 8.5$~\gevcc\ are
the resolutions.  Selected events have $r<2$.

\subsubsection{Neural Network Selection} 

\begin{figure}[t]
\begin{center} 
\epsfxsize=0.8\textwidth\epsfbox{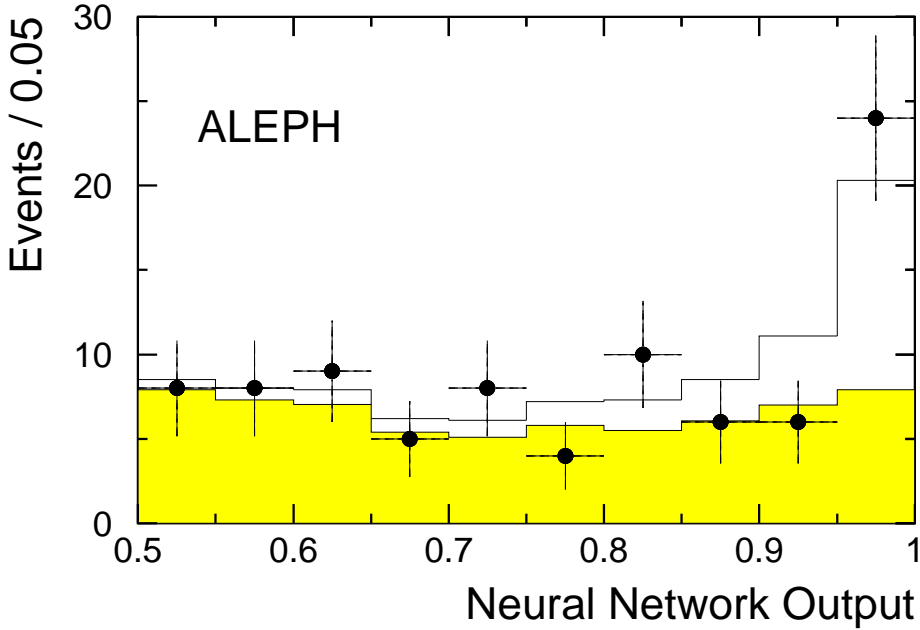}
\end{center}
  \caption{Neural network output from the \qqnn\ selection for data
  and Monte Carlo simulation.  The grey histogram shows the expected
  background, and the hollow histogram shows the expected
  \nctwo\ signal.\label{fig:nunuNN}}
\end{figure}

After preselection, a 12-variable neural network is employed to
distinguish signal events from background.  Four of the variables used
in the neural network analysis (acoplanarity, minijet energy,
direction of missing momentum, and energy within a $12^\circ$ cone
around the beam axis) are shared with the cut analysis.  Additionally,
the neural network analysis uses the reconstructed \zed\ boson mass
and the missing mass directly rather than using an elliptical mass
cut.  This analysis also includes two \bq-tagging variables although
only a small fraction of events have a \bbnn\ final state. In
addition, acollinearity, the visible energy beyond $30^\circ$ of the
beam axis, the energy within a $30^\circ$ wedge of the direction of
the missing momentum, and the $z$-component of the momentum are used
as variables.  The output of the neural network is shown in
Figure~\ref{fig:nunuNN} for Monte Carlo simulation and for data.  Only
events with a network output greater than 0.5 are used for the cross
section calculation.

\subsubsection{Systematic Uncertainties}

The results have been corrected for unsimulated accelerator background
events which increase the energy deposited near the beam axis; this
effect was studied using random triggers.  In addition, half of the
correction is used as an estimate of the systematic uncertainty.
Similarly, the jet kinematics are corrected to make them correspond
better to those observed in the data.  Again, half of this correction
is used as the systematic uncertainty.  The systematic uncertainties
also include uncertainties from the background production cross
sections and the limited statistics of the \MC\ samples.

Adding these uncertainties in quadrature gives total relative
uncertainties of 2\% and 5\% on the signal and background,
respectively, for the cut-based analysis.  For the efficiency, all of
the sources are significant; for the background, the limited \MC\
statistics dominate.

For the NN-based analysis, the \bq-tagging systematic uncertainties
are estimated by reweighting the data with a function obtained from
comparison of \zed-peak Monte Carlo simulation and data.  The relative
systematic errors on signal and background are 2\% and 3\%,
respectively.

\subsection{\boldmath Selection of $\zz\to\llnn$ Final States\unboldmath}

The preselection requires events to have exactly two identified
electrons or muons and no other reconstructed charged particles.
These leptons must have the same flavour and opposite charge.  The
fraction of the centre-of-mass energy deposited in the detector more
than $30^\circ$ from the beam axis \ecenn\ must satisfy the
requirement $0.4<\ecenn<0.6$.  The angle between the two lepton
directions must be less than $178^\circ$.  For electron pair events,
$49\pm 2$ events are expected and 43 observed; for muon pair events,
$13.2\pm 0.3$ events are expected and 10 observed.

Cuts on other kinematic quantities further suppress the background.
The invariant mass of the leptons and the missing mass are required to
be within an ellipse of $r<1.7$ defined as above with $\sigma_{\not
M}=3.3$~\gevcc\ and $\sigma_{m_{\ell\ell}}=2.5$~\gevcc.  The angle
between the missing momentum and the beam axis must be larger than
$6.7^\circ$, and the total energy not associated with the leptons must
be less than 5.6~\gev.

The systematic uncertainty is 5\% for the efficiency and 17\% for the
background.  These include uncertainties from lepton identification
and tracking resolution but are dominated by the limited \MC\
statistics.
   
\subsection{\boldmath Selection of $\zz\to\ttqq$ Final States \unboldmath}

In the neural network based selection, a dedicated analysis selects
$\zz\to \ttqq$ events with a higher efficiency than the
cut-based four-jet selection.

Events must contain at least eight tracks and the total energy of all
the charged particles must exceed $0.20\,\roots$.  Background from
\ww\ and \zz\ events with electron or muon decays is suppressed by
rejecting events having an identified lepton with energy greater than
$0.25\,\roots$.  Radiative returns to the \zed\ peak, characterized by
high missing energy and high missing longitudinal momentum, are
rejected by requiring $\left| \pzvis\right| + \emis < 1.8\,\eisr$,
where \pzvis\ and \emis\ are respectively the missing longitudinal
momentum and missing energy.  To further reject radiative returns,
events are also rejected if the event $\left|\pzvis\right|$ is greater
than $0.6\,\eisr$.

Events passing the preselection cuts are clustered into minijets
having invariant masses consistent with a tau lepton, $m_{\rm
jet}<2.7$~\gevcc.  The $\tau$ candidates are selected from these
minijets using a series of quality cuts based on multiplicity,
isolation, and momentum.  To be considered a $\tau$ candidate, a
minijet must have one, two, or three charged tracks with momenta
larger than 1~\gevc.  If the minijet has three charged tracks, it must
be of unit charge; if the minijet has two charged tracks, the minijet
charge is the charge of the track with higher momentum.  The minijet
isolation angle must be larger than $15^\circ$.  This isolation angle
is the half-angle of the largest cone around the minijet direction
containing, in addition to the minijet energy, no more than 5\% of the
total event energy outside the cone.  Finally, the energy of a two- or
three-prong minijet must be greater than 12.5~\gev, while a one-prong
minijet with less than 80\% of its energy from charged particles must
have an energy greater than 7.5~\gev.  Otherwise, no momentum cut is
made.

Only events with at least two $\tau$ candidates are further
considered.  At least one of the $\tau$ minijets must have exactly one
prong, and the two minijets must have opposite charge.  The rest of
the event is clustered into two jets using the Durham algorithm.  All
four jets in the event are rescaled using an overconstrained kinematic
consistency fit in which the jet directions are fixed and the minijet
masses are set to $m_\tau$.  The fit $\chi^2$ is calculated from
energy-momentum conservation, the hadronic jet resolutions, and the
difference between the two fitted dijet masses.  This has the effect
of constraining the dijets to equal masses.  In no case are the
hadronic jets allowed to rescale to less than $75\%$ of their measured
momenta.  A typical event may have several possible combinations of
potential $\tau$ minijet candidates; only the combination with the
smallest calculated kinematic $\chi^2$ is further considered.

A five-variable neural network selects \zz\ events from preselected
events.  Because the $\tau$ leptons decay with one or two neutrinos,
the event transverse momentum \ptvis\ is input to discriminate against
background events without missing energy.  The $\chi^2$ provides a
measure of the event's kinematic consistency with the \zz\ signal. The
sum of the $\tau$ candidates' isolation angles and the sum of the
$\tau$ candidates' transverse momenta with respect to their nearest
hadronic jet distinguish $\tau$ minijets, which are expected to be
well-isolated, from fake $\tau$ minijets on the periphery of a
hadronic jet.  Finally, the reconstructed \zed\ mass, when combined
with the implicit equal mass constraint, helps discriminate between
\nctwo\ events and background.  Events with a network output greater
than 0.77 are selected for the cross section calculation.

The systematic uncertainties arise from uncertainties on the
reconstructed jet energies and angles, the background cross sections,
and the limited \MC\ statistics.  The relative systematic errors for
signal and background are 3.0\% and 10.8\%.  The error for signal is
distributed equally among the sources, while the background error is
dominated by limited \MC\ statistics.

\section{Combination of Channels}

Table~\ref{tab:summary}\ summarizes the efficiencies, numbers of
expected signal and background events, the number of observed events,
and the measured cross section for each channel.  A maximum likelihood
fit determines the \nctwo\ cross section for the $\ee\to\zz$ process
by combining the information from all of the channels.  

For the cut-based analysis, only the total numbers of events are used
in the likelihood combination. Overlaps between the various channels
are less than 0.2\% and are negligible.  

For the given signal and background numbers, toy simulated experiments
are generated, and the expected uncertainty on the cross section
measurement is derived from the width of the extracted cross section
distribution. The expected relative statistical uncertainty on the
\elepB~\gev\ cross section measurement is \xsreB\%.

For the neural-network based analysis, a binned likelihood fit is used
for the \qqqq\ and \qqnn\ channels where the shape of the
neural-network distribution is used; for the other channels only the
total numbers of events are used in the likelihood fit.  The \qqqq\
distribution is divided into six bins between 0.7 and 1.0, while the
\qqnn\ distribution is divided into ten bins between 0.5 and 1.0.

For the neural-network based analysis, the overlaps between channels
are explicitly removed. About 0.2\% of the events in the \qqqq\
selection are \ttqq\ events in which a three-prong $\tau$ is
misidentified as a quark jet, and about 0.5\% are \qq\mm\ events.  To
ensure the exclusivity of the analyses, the overlapping signal and
background are explicitly subtracted from the \qqqq\ results.  The
expected uncertainty after combining all channels is \nxsres\%.

The systematic uncertainties were determined by adding a Gaussian
smearing to the efficiencies and background estimates in toy \MC\
experiments and observing the change in the total error.  Correlated
contributions to the total systematic uncertainty of the measurement
include the luminosity (0.5\%) and the uncertainties in the \ww\ and
\qq\ cross sections.  The total relative systematic uncertainties are
4\% and 6\% for the cut-based and neural-network based analyses,
respectively.

The measured cross sections at $\roots=\elepB$~\gev\ are
$$\sigma_{\zz} = \xsB \pm \xsstB\,\stat \pm \xssyB\,\syst\, \pb$$
and $$\sigma_{\zz} = \nxsec \pm \nxsstat\,\stat \pm
\nxssys\,\syst\, \pb$$ for the cut-based and neural-network based
analyses, respectively, compared to the \SM\ expectation of
\smxsB~\pb.  Table~\ref{tab:summary}\ shows the cross section for each
channel individually.

The cut-based analysis selects 13 candidates with an expected
background of 12.2 events when applied to the 1997 data sample.  The
total efficiency for all channels is 33\%.  Combination of the
channels with the above likelihood function yields a measured cross
section of
$$\sigma_{\zz} = \xsA \pm \xsstA\,\stat \pm \xssyA\,\syst\, \pb
$$ at a centre-of-mass energy of $\roots=\elepA$~\gev, compared to the
\SM\ expectation of \smxsA~\pb.  The measurement is also consistent
with a similar measurement from the L3 collaboration~\cite{Lthree}.

\begin{figure}[t]
\begin{center}
\epsfxsize=0.8\textwidth\epsfbox{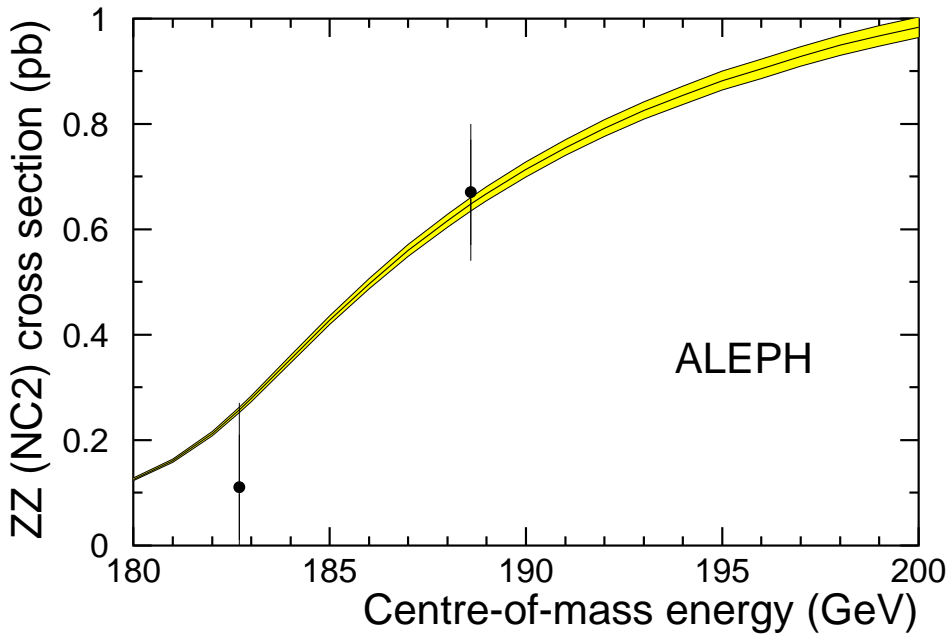}
\end{center}
\caption{The measured \nctwo\ cross section as a function of the
  centre-of-mass energy compared to the expected \SM\ cross section.
  The shaded area represents the theoretical uncertainty (2\%)
  on the \yfszz\ calculations.\label{fig:results}}
\end{figure}

\section{Cross Checks\label{xcheck}}

\subsection{Four-fermion Interference\label{xcheck:a}}

The \pythia\ \MC\ generator does not fully simulate the interference
between all of the four-fermion diagrams.  This potentially biases the
cross section measurement presented here.  Using \excalibur, which
does include the interference, to generate samples for the signal and
four-fermion background and rerunning the cut-based analysis yields a
relative difference in the cross section measurement of $-1\%$.  This
is a negligible effect when compared to the magnitude of the
statistical error.

\subsection{\boldmath Correlations Between the \ww\ and \zz\ Cross
Sections \unboldmath\label{xcheck:b}}

The preliminary measurement of the \ww\ cross section in the \qq\qq\
channel performed by \ALEPH\ on the 1998 data differs from the \SM\ by
$-10\%$.  Although this could be a statistical effect, the effect of a
real reduction of the \ww\ cross section on the extracted \zz\ cross
section was investigated.  As a check, the expected \ww\ background of
the four-quark channels of the cut-based analysis was reduced by 10\%
and the channels recombined.  This leads to a $+3.5\%$ relative change
in the measured \zz\ cross section.





\section{Conclusions}

Two analyses have measured the \zz\ cross section using the data taken
at $\roots=\elepB$~\gev\ in 1998\@.  The results agree and are
consistent with the \SM\ expectation of \smxsB~\pb.  The \zz\
production cross section is taken to be the arithmetic average of the
above two measurements
$$\sigma_{\zz}(\elepB~\gev) = \xsave \pm \xsavest\,\stat \pm
\xsavesy\,\syst\, \pb\,.$$
Additionally, the cut-based analysis has
been applied to the 1997 data sample yielding a measurement of
$$\sigma_{\zz}(\elepA~\gev) = \xsA \pm \xsstA\,\stat \pm
\xssyA\,\syst\, \pb \,.$$ The measured cross section is compared to
the \SM\ expectation in Figure~\ref{fig:results}.


\section*{Acknowledgements}

The authors would like to thank W.~P\l aczek for his help with the
\yfszz\ generator.  It is also a pleasure to congratulate our
colleagues from the \CERN\ accelerator divisions for the successful
operation of \LEP\@.  We are indebted to the engineers and technicians
in all our institutions for their contributions to the excellent
performance of \ALEPH.  Those of us from non-member states thank
\CERN\ for its hospitality.

\end{document}